\documentclass{elsart5p}

\usepackage{graphicx}

\usepackage{amssymb}
\usepackage{amsmath}

\begin{document}

\begin{frontmatter}

\title{Magnetic polaron structures in the one-dimensional double and super-exchange model}

\author[aff1]{E. Vallejo\corauthref{cor1}}
\ead{emapion@yahoo.com}
\corauth[cor1]{Corresponding author.}
\author[aff2]{F. L\'{o}pez-Ur\'{\i}as}
\author[aff1]{O. Navarro}
\author[aff3]{and M. Avignon}
\address[aff1]{Instituto de Investigaciones en Materiales, Universidad Nacional Aut\'{o}%
noma de M\'{e}xico, Apartado Postal 70-360, 04510 M\'{e}xico D. F., M\'{e}%
xico.}
\address[aff2]{Advanced Materials Department, IPICYT, Camino a la Presa San Jos\'{e} 2055,
Lomas 4a secci\'{o}n, 78216, San Luis Potos\'{\i}, San Luis Potos\'{\i}, M%
\'{e}xico.}
\address[aff3]{Institut N\'{e}el, Centre National de la Recherche Scientifique (CNRS) and
Universit\'{e} Joseph Fourier, BP\ 166, 38042\ Grenoble Cedex 9, France.}



\begin{abstract}
An analytical and numerical study of the one-dimensional double and
super-exchange model is presented. A phase separation between ferromagnetic
and anti-ferromagnetic phases occurs at low super-exchange interaction
energy. When the super-exchange interaction energy gets larger, the
conduction electrons are self-trapped within separate small magnetic
polarons. These magnetic polarons contain a single electron inside two or
three sites depending on the conduction electron density and form a Wigner
crystallization. A new phase separation is found between these small
polarons and the anti-ferromagnetic phase. Our results could explain the
spin-glass-like behavior observed in the nickelate one-dimensional compound $%
Y_{2-x}Ca_{x}BaNiO_{5}$.
\end{abstract}


\begin{keyword}
Exchange and super-exchange interactions\sep Classical spin models \sep Phase separation
\PACS 75.30.Et\sep 75.10.Hk\sep 64.75.+g
\end{keyword}
\end{frontmatter}

\section{Introduction}\label{}

Magnetic ordering of localized spins mediated by nonmagnetic conduction
electrons, the so-called double exchange (DE) or indirect exchange, is the
source of a variety of magnetic behavior in transition metal and rare-earth
compounds \cite{Mattis2006}. Conversely, this interplay affects the mobility
of the carriers and may lead to interesting transport properties such as
colossal magnetoresistance in manganites. The origin of the DE lies in the
intra-atomic coupling of the spin of the itinerant electrons with localized
spins $\overrightarrow{S}_{i}$. In this coupling, localized and itinerant
electrons belong to the same atomic shell.\ According to Hund's rule, the
coupling is ferromagnetic (F) when the local spins have less than
half-filled shells and anti-ferromagnetic (AF) for more than half-filled
shells \cite{anderson1955}. This mechanism has been widely used in the
context of manganites \cite{anderson1955,jonker1950,zener1951}. A similar
coupling also occurs in Kondo systems via the so-called $s-d$ exchange
model. In this case, local spins are from a $d$ shell (or $f$ shell in
rare-earth compounds) while the conduction electrons are from $s$ or $p$
states and the coupling is anti-ferromagnetic. In recent literature the
ferromagnetic coupling case is often referred to as the Ferromagnetic Kondo
model. Independently of the sign of the coupling, the \textquotedblleft
kinetic" energy lowering, favors a F background of local spins. This F
tendency is expected to be thwarted by AF super-exchange (SE) interactions
between localized spins $\overrightarrow{S}_{i}$ as first discussed by de
Gennes \cite{degennes1960} who conjectured the existence of canted states.\
In spite of recent interesting advances, our knowledge of magnetic ordering
resulting from this competition is still incomplete.

Although it may look academic, the one-dimensional (1D) version of this
model is very illustrative and helpful in building an unifying picture. On
the other hand, the number of pertinent real 1D systems as the nickelate
one-dimensional metal oxide carrier-doped compound $Y_{2-x}Ca_{x}BaNiO_{5}$%
\cite{DiTusaKojima} is increasing. In this compound, carriers are
essentially constrained to move parallel to $NiO$ chains and a
spin-glass-like behavior was found at very low temperature $T\lesssim 3K$
for typical dopings $x=0.045$, $0.095$ and $0.149$. Recently, it has been shown
that three-leg ladders in the oxyborate system Fe$_{3}$BO$_{5}$ may provide
evidence for the existence of spin and charge ordering resulting from such a
competition \cite{vallejo2006}.\newline

Naturally, the strength of the magnetic interactions depends significantly
on the conduction band filling, $x$. At low conduction electron density, F
polarons have been found for localized $S=1/2$ quantum spins \cite%
{batista19982000}. \textquotedblleft Island\textquotedblright\ phases,
periodic arrangement of F polarons coupled anti-ferromagnetically, have been
clearly identified at commensurate fillings both for quantum spins in one
dimension \cite{garcia20002002} and for classical spins in one \cite%
{koshibae1999} and two dimensions \cite{aliaga2001}. Phase separation
between hole-undoped anti-ferromagnetic and hole-rich ferromagnetic domains
has been obtained in the Ferromagnetic Kondo model \cite{yunokidagotto1998}.
Phase separation and small ferromagnetic polarons have been also identified
for localized $S=3/2$ quantum spins \cite{neuber2006}. Therefore, it is of
importance to clarify the size of the polarons, and whether it is preferable
to have island phases, separate small polarons or eventually large polarons.%
\newline

In this paper, we present an analytical and numerical study of the
one-dimensional double and super-exchange model. Our results provide a
plausible understanding of the ground state of the nickelate 1D compound $Y_{2-x}Ca_{x}BaNiO_{5}$
allowing a straightforward explanation of its spin-glass-like behavior \cite%
{DiTusaKojima}. The paper is organized as follows. In section 2 a brief
description of the model is given. In section 3, results and a discussion
are presented. Finally, our results are summarized in section 4.

\section{The model}

The DE\ Hamiltonian is originally of the form, 
\begin{equation}
H=-\sum_{i,j;\sigma }t_{ij}(c_{i\sigma }^{+}c_{j\sigma }+h.c.)-J_{H}\sum_{i}%
\overset{\rightarrow }{S_{i}}\cdot \overset{\rightarrow }{\sigma }_{i},
\label{A}
\end{equation}%
where $c_{i\sigma }^{+}(c_{i\sigma })$ are the fermions creation
(annihilation) operators of the conduction electrons at site $i$, $t_{ij}$
is the hopping parameter and $\overrightarrow{\sigma }_{i}$ is the
electronic conduction band spin operator. In the second term, $J_{H}$ is the
Hund's exchange coupling. Here, Hund's exchange coupling is an intra-atomic
exchange coupling between the spin of conduction electrons $\overrightarrow{%
\sigma }_{i}$ and the spin of localized electrons $\overrightarrow{S}_{i}$.
This Hamiltonian simplifies in the strong coupling limit $J_{H}\rightarrow
\infty $, a limit commonly called itself the DE model. We will consider the
local spins as classical $\overrightarrow{S}_{i}$ $\rightarrow \infty $, a
reasonable approximation in many cases in view of the similarity of the
known results \cite{garcia20002002,yunokidagotto1998}. The DE Hamiltonian
takes the well-known form, 
\begin{equation}
H=-\sum_{i,j}t_{i,j}\cos \left( \frac{\theta _{i,j}}{2}\right)
(c_{i}^{+}c_{j}+h.c.).  \label{B}
\end{equation}%
The itinerant electrons being now either parallel or antiparallel to the
local spins are thus spinless.\ $\theta _{i,j}$ is the relative angle
between the classical localized spins at sites $i$ and $j$ which are
specified by their polar angles $\phi _{i}$, $\varphi _{i}$ defined with
respect to a $z$-axis taken as the spin quantization axis of the itinerant
electrons. The super-exchange coupling is an anti-ferromagnetic inter-atomic
exchange coupling between localized spins $\overrightarrow{S}_{i}$. The
complete one-dimensional DE+SE Hamiltonian becomes, 
\begin{equation}
H=-t\sum_{i}\cos \left( \frac{\theta _{i}}{2}\right)
(c_{i}^{+}c_{i+1}+h.c.)+J\sum_{i}\overset{\rightarrow }{S}_{i}\cdot \overset{%
\rightarrow }{S}_{i+1},  \label{C}
\end{equation}%
where $\theta _{i,i+1}=\theta _{i}$ and $J$ is the super-exchange
interaction energy.

\section{Results and discussion}

In this section, we determine the complete phase diagram in one dimension as
a function of the super-exchange interaction energy $J$ and the conduction
electron density $x$, showing that up to now the model has not revealed all
its richness. Besides the quantum results already published \cite%
{batista19982000,garcia20002002,neuber2006} we find two types of phase
separation. In addition to the expected F-AF phase separation appearing for
small $J$, we obtain a new phase separation between small polarons (one
electron within two or three sites) and AF regions for larger $J$. It is
interesting to note that large polarons are never found stable in this limit.

The magnetic phase diagram has been obtained at T=0K by using open
boundary conditions on a linear chain of N=60 sites. For a given
conduction electron density $x$ ($0\leq x\leq 0.5$ because of the
hole-electron symmetry), we have to optimize all the N-1 angles $\theta
_{i}$. For this goal, we use an analytical optimization and a classical
Monte Carlo method. The analytical solution has been tested as a starting
point in the Monte Carlo simulation.

\begin{figure}[h]
\centering  \includegraphics[scale=0.8]{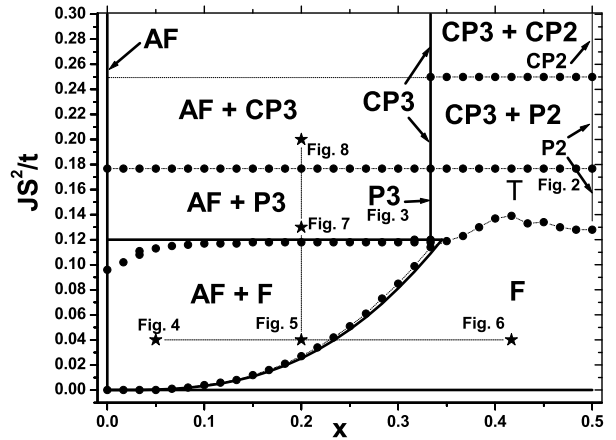}
\caption{Magnetic phase diagram as a function of the SE energy $J$ and the
conduction electron density $x$. A dotted line in this diagram represents a
guide for the eyes. The different phases are described in the text.}
\label{Fig1CL}
\end{figure}

Our results are summarized in figure \ref{Fig1CL}, showing the whole
magnetic phase diagram. 

For the commensurate fillings $x=1/2$ and $1/3$, we recover the
\textquotedblleft island" phases with ferromagnetic polarons ($\theta _{i}=0$%
) separated by antiferromagnetic links ($\theta _{i}=\pi $), P2 ($\cdots
\uparrow \uparrow \downarrow \downarrow \uparrow \uparrow \downarrow
\downarrow \uparrow \uparrow \downarrow \downarrow \cdots $) and P3 ($%
\cdots \uparrow \uparrow \uparrow \downarrow \downarrow \downarrow \uparrow
\uparrow \uparrow \downarrow \downarrow \downarrow \cdots $), figure \ref%
{Fig7CL} and figure \ref{Fig8CL} respectively, identified previously for
classical \cite{koshibae1999} and $S=1/2$ quantum \cite{garcia20002002} local
spins. In the quantum case, the real space spin-spin correlations illustrate
such structures. For these phases, the analytical optimization implies
angles $0$ or $\pi $ exactly.

\begin{figure}[h]
\centering  \includegraphics[scale=0.8]{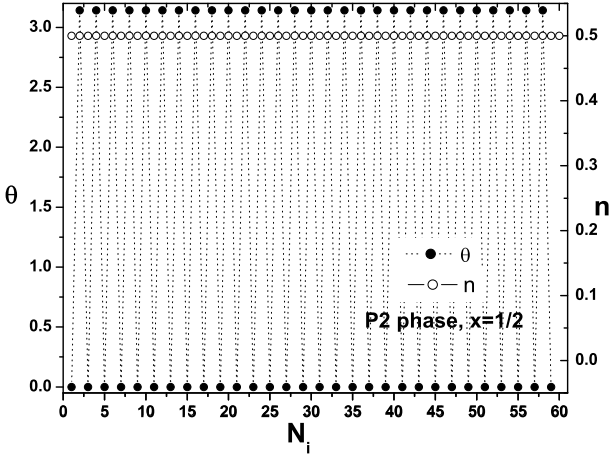}
\caption{P2 phase for $x=1/2$, showing N-1 angles ($\theta$) and charge distribution (n).
Angles in this figure are $0$ or $\protect\pi$ exactly.}
\label{Fig7CL}
\end{figure}

\begin{figure}[h]
\centering  \includegraphics[scale=0.8]{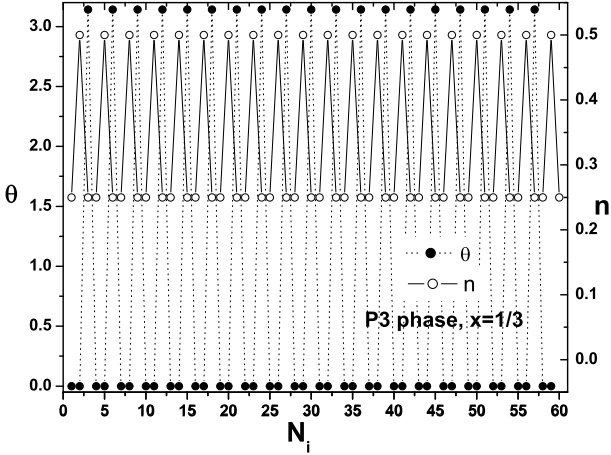}
\caption{P3 phase for $x=1/3$, showing the same as in figure \ref{Fig7CL}.}
\label{Fig8CL}
\end{figure}

The electrons are individually self-trapped in small independent
ferromagnetic polarons of two and three sites respectively forming a Wigner
crystallization.  In reference \cite{yunoki1998}, a spiral phase has been
proposed instead of the P2 phase for $x=1/2$. The ferromagnetic phase is
stable for weak SE interaction below P2 phase, P2 phase becomes stable
for $\frac{2}{\pi }-\frac{1}{2}<JS^{2}/t<\frac{1}{4}$. For $JS^{2}/t>\frac{1%
}{4}$, P2 transforms into a canted polaron phase CP2 in which the angle
inside the F islands becomes finite ($\theta _{1}$) while the angle between
the polarons ($\theta _{2}$) still keeps the value $\pi $. A complete
analytical solution can be derived in this case.

\begin{figure}[t]
\centering  \includegraphics[scale=0.8]{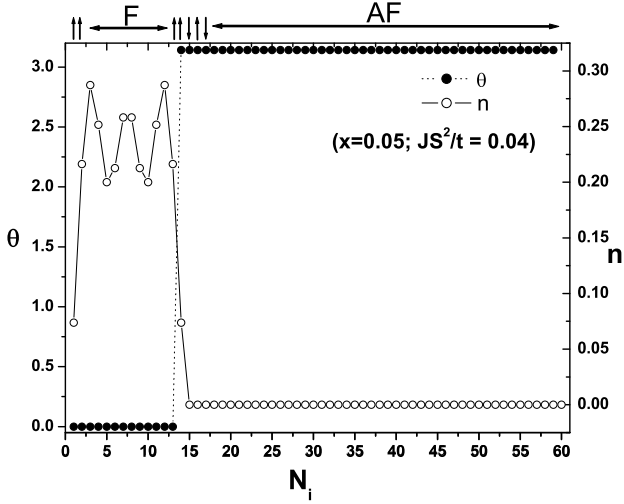}
\caption{AF+F phase at $x=0.05$ (3 electrons) and $JS^{2}/t=0.04$, showing N-1 angles, 
charge distribution and a spin
configuration snapshot.}
\label{Fig2CL}
\end{figure}

\begin{figure}[h]
\centering  \includegraphics[scale=0.8]{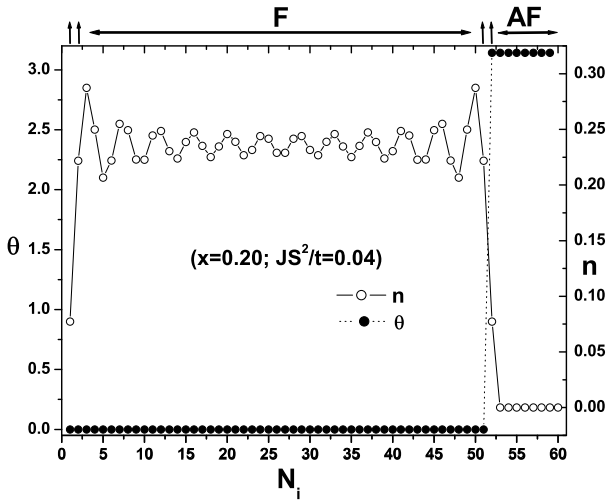}
\caption{The same as in Fig. \protect\ref{Fig2CL}, but at $x=0.20$ (12
electrons). }
\label{Fig3CL}
\end{figure}

\begin{figure}[h]
\centering  \includegraphics[scale=0.8]{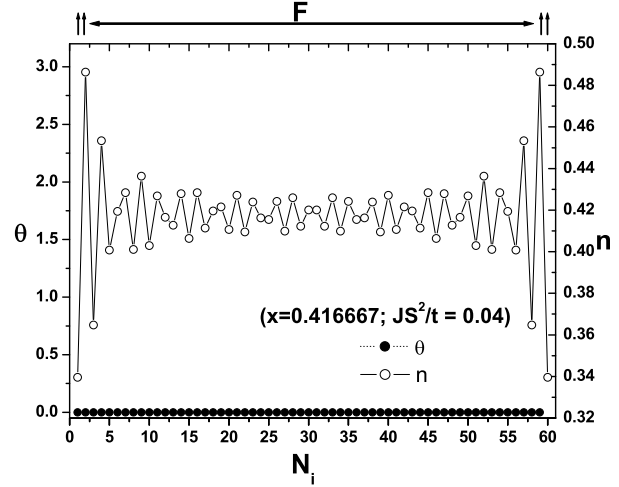}
\caption{F phase at $x=\frac{25}{60}$ (25 electrons) and $JS^{2}/t=0.04$,
showing the same as in figure \ref{Fig2CL}.}
\label{Fig4CL}
\end{figure}

\begin{figure}[t]
\centering  \includegraphics[scale=0.8]{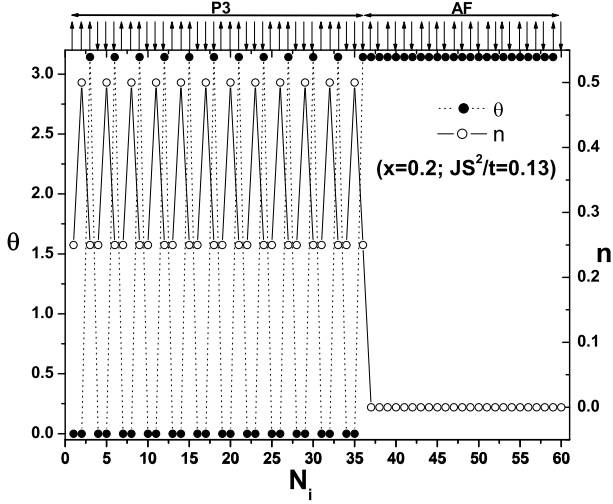}
\caption{AF+P3 phase at $x=0.20$ (12 electrons) and $JS^{2}/t=0.13$, showing 
N-1 angles, charge distribution and a spin
configuration snapshot.}
\label{Fig5CL}
\end{figure}

\begin{figure}[h]
\centering  \includegraphics[scale=0.8]{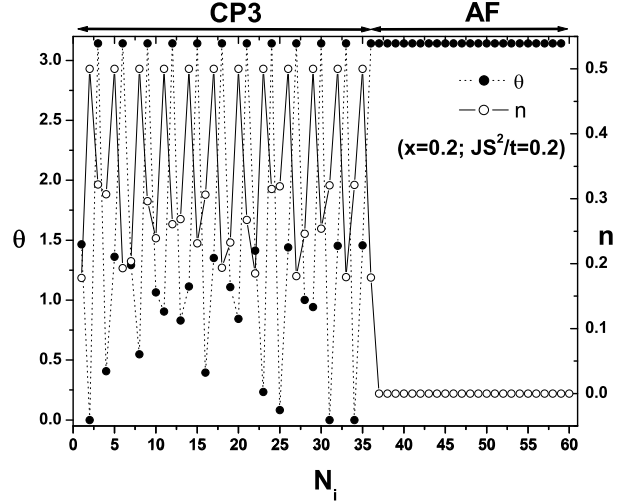}
\caption{AF+CP3 phase at $x=0.20$ (12 electrons) and $JS^{2}/t=0.20$,
showing the same as in figure \ref{Fig5CL}.}
\label{Fig6CL}
\end{figure}

Similar phases P3 and CP3 are obtained for $x=1/3$. In CP3, two angles 
$\theta _{1}$, $\theta _{2}$ are finite inside the 3-site polaron while,
between polarons, $\theta _{3}=\pi $. This phase has a general continuous
degeneracy within each 3-site polaron given by the condition, 
\begin{equation}
\cos (\theta _{1})+\cos (\theta _{2})=\frac{1}{8(JS^{2}/t)^{2}}-2.  \label{D}
\end{equation}%
An example of this degeneracy of the spin configuration is clearly seen in
fig \ref{Fig6CL} where different sets of angles ($\theta _{1},\theta _{2}$)
appear within the CP3 phase. Both CP2 and CP3 evolve towards complete
anti-ferromagnetism as $JS^{2}/t\rightarrow \infty $, see equation (\ref{D}). 
$P3\rightarrow CP3$ \ at $\frac{JS^{2}}{t}=\frac{1}{4\sqrt{2}}.$ These
phases result from the \textquotedblleft spin-induced Peierls $2k_{F}$
instability\textquotedblright\ due to the modulation of the hopping with $%
I=1/x$ angles. For lower commensurate fillings $x<1/3$, such $P_{I}$ polaron
phases are not found stable. Instead, next to the F phase at low $J$ we find
AF-F phase separation. Of course, an anti-ferromagnetic phase always occurs
at $x=0$. Figure \ref{Fig1CL} shows that when the SE interaction energy is
small $JS^{2}/t\lesssim 0.12$, the F phase occurs for a large conduction
electron density.\newline
The F-AF transition is given by the F-AF phase separation (AF+F in figure %
\ref{Fig1CL}) consisting of one large ferromagnetic polaron within an AF
background as can be seen in figures \ref{Fig2CL} and \ref{Fig3CL}, for a
typical value of $JS^{2}/t=0.04$. All electrons are inside the polaron. The
position of the polaron within the linear chain is not important because of
translation degeneracy. These figures also show charge distribution (n)
inside each polaron and a spin configuration snapshot. In this region, the
polarons' size diminishes with the conduction electron density, (figures \ref%
{Fig2CL} and \ref{Fig3CL}). For this F-AF phase separation, the analytical
optimization implies angles $0$ and $\pi $ exactly for the F and AF domains
respectively (figures \ref{Fig2CL} and \ref{Fig3CL}). In the thermodynamic
limit $N\rightarrow \infty ,$ and for $M\gg 3$ sites ($M$ being the size of
the F domain), the energy is obtained as, 
\begin{equation}
\frac{U}{Nt}=-2x\cos \left( x_{o}\pi \right) -\frac{JS^{2}}{t},  \label{F}
\end{equation}%
for $x\leq x_{o}.$ $x_{o}$ corresponds to the Maxwell construction between
the anti-ferromagnetic energy $\frac{U}{Nt}=-\frac{JS^{2}}{t}$ and the
ferromagnetic one $\frac{U}{Nt}=-\frac{2}{\pi }\sin \left( x\pi \right) +%
\frac{JS^{2}}{t}.$ $x_{o}$ is given by the following equation, 
\begin{equation}
x_{o}\cos \left( x_{o}\pi \right) -\frac{1}{\pi }\sin \left( x_{o}\pi
\right) +\frac{JS^{2}}{t}=0.  \label{G}
\end{equation}%
The corresponding boundary given by equation (\ref{G}) is shown by the full
line in figure \ref{Fig1CL}. The analytical results for N=60 sites are
very close to this line. When the conduction electron density gets larger $%
x\geq x_{o},$ the F-AF phase separation becomes the F phase. The size of the
ferromagnetic polaron in the thermodynamic limit $(N\rightarrow \infty ,M\gg
3)$ is $\epsilon =\frac{M}{N}=\frac{x}{x_{o}},$ for $x\leq x_{o},$ and $%
\epsilon =1$ in the ferromagnetic phase. An important effect of lattice
distortion is expected between these F and AF domains \cite{vallejo2006}.
The effect of lattice distortion can be studied in the F-AF phase separation
using the following density matrix elements. Inside the F domain the matrix
elements are given by, 
\begin{equation}
\rho _{i,j}=\frac{2}{M+1}\sum_{p=1}^{xN}\left( \sin \frac{(p)(i)\pi }{M+1}%
\right) \left( \sin \frac{(p)(j)\pi }{M+1}\right) .  \label{H}
\end{equation}%
Between F-AF domains and within AF domains the matrix elements are zero.
These density matrix elements suggest an important lattice distortion inside
F domains and null between F-AF domains and within AF domains. This lattice
distortion could be detectable for example using neutron diffraction
techniques as in $La_{2}CuO_{4+\delta }$ \cite{yunokidagotto1998}. Charge
distribution is easily obtained for $n_{i}=\rho _{i,j=i}$ figures \ref%
{Fig7CL}-\ref{Fig5CL}. For example in figure \ref{Fig4CL}, charge
distribution of the F phase can be observed for 25 electrons and at $\frac{%
JS^{2}}{t}=0.04.$ 
For small SE interaction, the F-AF phase separation has been reported in two
dimensions \cite{yamanaka1998}, in one dimension using classical localized
spins and $J_{H}=8$ \cite{yunoki1998} and in the one-dimensional
ferromagnetic Kondo model \cite{koller2003}. Quantum results for $S=3/2$,
showed phase separation when Coulomb repulsion was taken into account \cite%
{neuber2006}. We can see that in this limit, our results differ from those
of Koshibae \textit{et al.} \cite{koshibae1999} within the \textquotedblleft
spin-induced Peierls instability\textquotedblright\ mechanism.\newline

At low concentration $x<1/3$, if the SE interaction energy increases $0.12\lesssim $ $JS^{2}/t$ $\lesssim \frac{1}{4\sqrt{2}}$,
we find a new phase separation between P3 and AF phases as shown in figures \ref{Fig1CL} and \ref%
{Fig5CL}. It transforms into AF+CP3 for $JS^{2}/t$ $>\frac{1}{4\sqrt{2}}$ as P3
becomes CP3. A phase like AF+P3(CP3) has been identified using $%
S=3/2$ quantum spins \cite{neuber2006}. Figure \ref{Fig5CL} and figure \ref%
{Fig6CL} show the AF+P3 and the AF+CP3 phases with $12$ electrons among
the $60$ sites for typical values of the SE interaction energy $JS^{2}/t=0.13
$ and $JS^{2}/t=0.20$ respectively. These phase separations consisting in P3 or CP3
phases in an AF background and they are degenerate with phases where the
polarons can be ordered or not, while keeping the number of F and AF bonds
fixed. The phase obtained within the \textquotedblleft spin-induced Peierls
instability\textquotedblright\ \cite{koshibae1999} belongs to this class.
The former degeneracy unifies ideas like phase separation and individual
polarons and gives a natural response to the instability at the Fermi energy
and to an infinite compressibility as well. In the thermodynamic limit,
P3-AF phase separation energy is given by the following equation 
\begin{equation}
\frac{U}{Nt}=\left( -\sqrt{2}+4\frac{JS^{2}}{t}\right) x-\frac{JS^{2}}{t}.
\label{I}
\end{equation}%
We find that the $AF+F\rightarrow AF+P3$ transition is first order. In
figure \ref{Fig1CL}, the transition line $JS^{2}/t\simeq 0.12$ between the
two phase separations AF+P3 and AF+F has been determined using the
corresponding energies in the thermodynamic limit. Density matrix elements
suggest a large lattice distortion within each 3-site polaron in P3 phase.
Charge distribution among the 3 sites $i=1,2,3$ inside each polaron is $%
n_{1}=n_{3}=\frac{1}{4}$ and $n_{2}=\frac{1}{2}$, see figure \ref{Fig5CL}.%
\newline

Phase separation also takes place for fillings between $x=1/2$
and $x=1/3$ for SE interactions $JS^{2}/t\geq \frac{1}{4\sqrt{2}}$. It is
between CP3 and P2 or CP2 due to the canting inside the P2 polaron
with increasing $J$. The transition between the two occurs for $JS^{2}/t=0.25
$, where $P2\rightarrow CP2$. Again, due to the AF links between the
polarons these phase separations are degenerate with respect to the position
of the two types of polarons. Below CP3+P2, the phase labelled T in
figure \ref{Fig1CL} is a more general complex phase obtained by the Monte
Carlo method and can be polaronic like or not. Close to the boundary with CP3+P2 
this phase resembles the P3+P2 phase separation, so as seen in
figure \ref{Fig1CL}, the transition line $JS^{2}/t=\frac{1}{4\sqrt{2}}$,
corresponding to $P3\rightarrow CP3$, is second order.\ This boundary also
extends in the region $x<1/3$ between AF+P3 and AF+CP3.  
For the SE interaction region $JS^{2}/t>\frac{1}{4}$ ($JS^{2}/t=\frac{1}{4}$
is shown by the short-dotted line in figure \ref{Fig1CL}), the spin
configurations ($\theta _{1},\pi $) or ($\pi ,\theta _{2}$), i.e a CP2
polaron plus an AF link, belong to all the possible degenerate configurations
of CP3 polarons as can be seen from equation (\ref{D}).\ This means that
in this region phase separation AF+CP3 may also contain a number of
two-sites canted polarons CP2. Similarly, this also occurs within CP3+CP2 phase. 
The total number of polarons remaining equal to the number of
electrons; we can label it as AF+CP3+CP2. A single energy is found in
the whole conduction electron density regime $\left( 0\leq x\leq 0.5\right) $%
. In the thermodynamic limit, it corresponds to, 
\begin{equation}
\frac{U}{Nt}=-\frac{x}{8\frac{JS^{2}}{t}}-\frac{JS^{2}}{t}.  \label{J}
\end{equation}
The energies of each CP3 and CP2 polarons are respectively $-\frac{1}{8%
\frac{JS^{2}}{t}}-\frac{2JS^{2}}{t}$ and $-\frac{1}{8\frac{JS^{2}}{t}}-\frac{%
JS^{2}}{t}$.

Let us mention that homogeneous spiral phases ($\theta _{i}=\theta $) could
be possible ground states. In the thermodynamic limit, these
phases can occur for $\frac{JS^{2}}{t}\geq \frac{\sin \pi x}{2\pi }$ and
have energy $\frac{U}{t}=-\frac{(\sin \pi x)^{2}}{2\pi ^{2}(\frac{JS^{2}}{t})%
}-\frac{JS^{2}}{t}$. However, our Monte Carlo results show that these are
never stable within the model used here. This can be proved analytically in
the thermodynamic limit using the expressions we have derived for the
different phases, except for the T-phase for which numerical results are
necessary. In two-dimension however, renewed interest in a spiral state
results from experiments indicating a spin glass behavior of high-T$_{c}$ La$%
_{2-x}$Sr$_{x}$CuO$_{4}$ at small doping \cite{raczkowski2006}.

Of course, all the phase separations involving CP3 (AF+CP3, CP3+P2, CP3+CP2) 
present the spin configuration degeneracy ($\theta _{1},\theta _{2}
$) of the CP3 polarons. This analytical continuous degeneracy is
consistent with a spin glass state. Therefore, we propose that the ground
state of  $Y_{2-x}Ca_{x}BaNiO_{5}$ for the studied hole doping $x<0.15$ \cite%
{DiTusaKojima} belongs to the AF+CP3 phase providing a plausible
explanation for the observed spin-glass-like behavior. It is interesting to
note that such a possibility of polarons immersed into an anti-ferromagnetic
background has been invoked by Xu et al.\cite{Xu2000} to fit their neutron data.
Finally, we remark that the size chosen for the linear chain N=60 sites
and the boundary conditions do not change the nature of the phases involved
in the phase diagram.\newline

\section{Conclusions}

In this work, we presented an unifying picture for the magnetic phase
diagram of the one-dimensional DE+SE model using large Hund's coupling and
classical localized spins. The solution is in general a) phase separation
between F and AF phases for low SE interaction energy and b) phase
separation between small polaronic and AF phases when the SE interaction is
large. In a large SE limit a Wigner crystallization and a spin-glass
behavior can be identified. A spin-glass behavior can be obtained under the
condition $\frac{JS^{2}}{t}\gtrsim \frac{1}{4\sqrt{2}}\approx 0.177$, when
CP3 phase exists and could explain the spin-glass-like behavior observed in
the nickelate one-dimensional doped compound $Y_{2-x}Ca_{x}BaNiO_{5}$. On
the other hand, density matrix elements suggest an important lattice
distortion in the phases involved in the phase diagram. \newline

\bigskip

\begin{center}
\textbf{Acknowledgment}
\end{center}

We want to acknowledge partial support from CONACyT Grant-57929 and
PAPIIT-IN108907 from UNAM. E.V. would like also to thank CONACyT and
DGAPA-UNAM for financial support. 

\bigskip

\begin{thebibliography}{00}

\bibitem{Mattis2006} 
D.\ C.\ Mattis, The theory of magnetism made simple, World Scientific, Singapore, 2006.

\bibitem{anderson1955} 
P.\ W.\ Anderson and H.\ Hasegawa, Phys.\ Rev. \textbf{100} (1955) 675.

\bibitem{jonker1950} 
G.\ H.\ Jonker and J.\ H.\ Van Santen, Physica \textbf{16} (1950) 337; J.\ H.\ Van Santen and G.\ H.\ Jonker, Physica \textbf{16} (1950) 599.

\bibitem{zener1951} 
C.\ Zener, Phys.\ Rev. \textbf{82} (1951) 403; C.\ Zener, Phys.\
Rev. \textbf{81} (1951) 440.

\bibitem{degennes1960} 
P.\ G.\ de Gennes, Phys.\ Rev. \textbf{118} (1960) 141.

\bibitem{DiTusaKojima} 
J.\ F.\ DiTusa \textit{et al.}, Phys.\ Rev.\ Lett. \textbf{73} (1994) 1857; K.\ Kojima \textit{et al.}, Phys.\
Rev.\ Lett. \textbf{74} (1995) 3471.

\bibitem{vallejo2006} 
E.\ Vallejo and M.\ Avignon, Phys.\ Rev.\ Lett. \textbf{97} (2006) 217203; E.\ Vallejo and M.\ Avignon, Rev.\ Mex.\ F\'{i}s. S \textbf{53} (2007) (7) 1-6; E.\ Vallejo and M.\ Avignon, J.\ Magn.\ Magn.\ Mat. \textbf{310} (2007) 1130.

\bibitem{batista19982000} 
C.\ D.\ Batista, J.\ Eroles, M.\ Avignon and B.\ Alascio, Phys.
Rev. B \textbf{58} (1998) R14689; C.\ D.\ Batista, J.\ Eroles, M.\ Avignon
and B.\ Alascio, Phys. Rev. B \textbf{62} (2000) 15047.

\bibitem{garcia20002002} 
D.\ J.\ Garcia \textit{et al.}, Phys.\ Rev.\ Lett. \textbf{85} (2000) 3720; D.\ J.\ Garcia \textit{%
et al.}, Phys. Rev. B \textbf{65} (2002) 134444.

\bibitem{koshibae1999} 
W.\ Koshibae, M.\ Yamanaka, M. Oshikawa and S.\ Maekawa, Phys.\
Rev.\ Lett. \textbf{82} (1999) 2119.

\bibitem{aliaga2001} 
H.\ Aliaga \textit{et al.},
Phys.\ Rev.\ B, \textbf{64} (2001) 024422.

\bibitem{yunokidagotto1998} 
S.\ Yunoki \textit{et al.}, Phys.\ Rev.\ Lett. \textbf{80} (1998) 845; E.\ Dagotto \textit{%
et al.}, Phys. Rev. B \textbf{58} (1998) 6414.

\bibitem{neuber2006} 
D.\ R.\ Neuber \textit{et al.}, Phys. Rev. B \textbf{73} (2006) 014401.

\bibitem{yunoki1998} 
S.\ Yunoki and A.\ Moreo, Phys. Rev. B \textbf{58} (1998) 6403.

\bibitem{yamanaka1998} 
M.\ Yamanaka, W.\ Koshibae and S.\ Maekawa, Phys.\ Rev.\ Lett. 
\textbf{81} (1998) 5604.

\bibitem{koller2003} 
W.\ Koller \textit{et al.}, Phys. Rev. B \textbf{67} (2003) 174418.

\bibitem{raczkowski2006} 
M.\ Raczkowski, R.\ Fr\'esard and A.\ M.\ Ole\'s, Europhys. Lett. \textbf{76} (2006) 128.

\bibitem{Xu2000} 
G.\ Xu \textit{et al.}, Science, New Series \textbf{289} (2000) 419.




\end{thebibliography}
\end{document}